\documentclass[12pt]{article}
\usepackage{latexsym}
\newcommand{\binco}[2]{\renewcommand{\arraystretch}{0.7}
  \left(\begin{array}{@{\hspace{1pt}}c@{\hspace{1pt}}}#1\\#2\end{array}\right)
  \renewcommand{\arraystretch}{1}}
\newcommand{\rud}{\rule[-8pt]{0pt}{22pt}}
\newcommand{\ru}{\rule[0pt]{0pt}{14pt}}
\newcommand{\rd}{\rule[-8pt]{0pt}{8pt}}
\newcommand{\zr}[1]{\mbox{\hspace*{#1em}}}
\newcommand{\NN}{\mbox{\zr{0.1}\rule{0.04em}{1.6ex}\zr{-0.05}{\sf N}}}
\newcommand{\NNs}{\mbox{\small \zr{0.1}\rule{0.04em}{1.6ex}\zr{-0.05}{\sf N}}}

\newtheorem{theorem}{Theorem}
\newtheorem{lemma}{Lemma}
\newtheorem{prop}{Proposition}

\begin{document}

\begin{center}
{\LARGE\bf The Entropy of Square-Free Words} \\[8mm]
{\large\sc 
  Michael Baake$^{1}$, \hspace*{1pt} Veit Elser$^{1,2}$
  \hspace*{1pt} and \hspace*{1pt} Uwe Grimm$^{3,\dagger}$} \\[4mm]
{\footnotesize \mbox{}\footnotemark[1]
 Institut f\"{u}r Theoretische Physik, Universit\"{a}t T\"{u}bingen,\\
 Auf der Morgenstelle 14, 72076 T\"{u}bingen, Germany} \\[2mm]
{\footnotesize \mbox{}\footnotemark[2]
 Laboratory of Atomic and Solid State Physics,\\
 Cornell University, Ithaca, NY 14853-2501, USA} \\[2mm]
{\footnotesize \mbox{}\footnotemark[3]
 Instituut voor Theoretische Fysica, Universiteit van Amsterdam, \\
 Valckenierstraat 65, 1018 XE Amsterdam, 
 The Netherlands}
 \renewcommand{\thefootnote}{\fnsymbol{footnote}}\footnotetext[2]{
 Address after April 1996: Institut f\"ur Physik, Technische Universit\"at, 
 D-09130 Chemnitz, Germany}
  \\[8mm]
\end{center}


\bigskip
\centerline{\bf Abstract}

\begin{quote}
{\small\sf 
Finite alphabets of at least three letters permit the construction
of square-free words of infinite length. We show that the entropy
density is strictly positive and derive reasonable lower and upper
bounds. Finally, we present an approximate formula which is
asymptotically exact with rapid convergence in the number of letters.
}
\end{quote}

\bigskip
\centerline{\bf R\'esum\'e}

\begin{quote}
{\small\sf Il est possible de construire des mots de longueur infinie
sans carr\'e sur un alphabet ayant au moins trois lettres.  Nous
d\'emontrons que l'entropie du langage des mots sans carr\'e sur un
tel alphabet est strictement positive et l'encadrons par des bornes
inf\'erieure et sup\'erieure raisonnables.  Enfin, nous donnons pour
l'entropie une expression approch\'ee qui est asymptotiquement
correcte et converge rapidement lorsque le nombre de lettres de
l'alphabet tend vers l'infini.}

\bigskip
{\small\bf Key Words:} \hspace*{5pt} 
{\footnotesize  Combinatorics on words, Entropy,
                Substitution rules, Random tilings}
\end{quote}



\renewcommand{\theequation}{\arabic{section}.\arabic{equation}}

\section{Introduction}
\setcounter{equation}{0}

As is well-known, deterministic rules (like substitution rules etc.)
can only result in tilings or discrete structures with vanishing {\em
entropy density} \cite{Berthe} -- the famous Penrose tiling is an
example of this phenomenon. This tiling and many other ones appear in
the description of so-called {\em quasicrystals} \cite{Janot} for very
good reason: Locally finite discrete structures (such as tilings with
only finitely many prototiles) provide useful cell models for the
description of the ordered state, not only crystalline but also
quasi-crystalline \cite{Nissen}. In the latter case, one is
particularly interested in models with {\em finite} (i.e.,
non-vanishing) entropy density, as this is one possible mechanism to
explain non-periodic order through entropic stabilization.  In order
to combine long-range order with a decent amount of randomness,
so-called {\em random tiling models} have been studied in quite some
detail \cite{Henley}, and there exists a reasonable qualitative
understanding.

In view of these remarks it is clear that exactly solvable random
tiling models are of interest. The standard case in one dimension is
trivial as it is essentially equivalent to a Bernoulli scheme (cf.\
Ref.~\cite{Petersen}). This can hardly be called ordered in any sense.
In two dimensions, we know of only a few solved cases (such as the
Fisher-Kasteleyn domino model (see app.~E in Ref.~\cite{Thompson}),
the hexagonal random tiling model \cite{Henley} or the square-triangle
random tiling model \cite{Kalugin}), and many attempts to improve this
situation have failed so far, and the situation in higher dimensions
is even worse.  So the question arises whether one can find other
examples in 1D that are more restrictive than the Bernoulli scheme but
still provide reasonable toy models of (partially) ordered
states. Here one can scan the vast number of examples of automata and
other sequences \cite{Allouche}, but hardly any of them yield an
interesting model with positive entropy density.

One interesting class, however, is provided by infinite words in a
finite alphabet that avoid the repetition of certain patterns. The
simplest such case is the ensemble of {\em square-free words}, first
studied by Thue \cite{Thue1,Thue2} and later on reinvestigated many
times, see e.g.\ Refs.~\cite{Pleasants70}--\cite{Kobayashi}.  In
particular, the combinatorial problems in the treatment of these
systems are very interesting.  It turned out in a series of
independent articles \cite{Elser,Brandenburg,Brinkhuis} that the
entropy density of square-free words in three letters is {\em
positive}.  It was conjectured that the existing upper bounds were
much closer to the actual value of the entropy density than the lower
bounds, and that it is close to $0.3$.  We will see later that the
numerical value is in fact about $0.263719$.

In this article, we will summarize some of the properties of the
ensemble of square-free words in an alphabet $\cal A$ with finitely
many letters, $x$ say. (For a general background, we refer to
Ref.~\cite{Lothaire}, although we shall use slightly different
notation here as a compromise between mathematical and physical
literature.)  We will concentrate on the case $x=3$ for a while before
we treat the general case. We present various rigorous results but
also include new numerical calculations which finally guide us to the
conjecture of an asymptotically correct formula for the entropy
density of square-free words in $x$ letters which turns out to be
amazingly accurate already for small $x$.

\section{Basic setup and inequalities}
\setcounter{equation}{0}

Let $A=\{a_1^{},a_2^{}, \ldots , a_x^{}\}$ be a finite alphabet with
$x$ different letters (so $x \in \NN$ is an integer). Then, by ${\cal
A} = A^{\NNs_0}$, we denote the set of all words of finite length in
the letters of $A$. This is a monoid with concatenation of words as
operation and the empty word as neutral element \cite{Lothaire}.
Later on, we will restrict ourselves to subsets of ${\cal A}$ which
are more interesting. If we write $\ell(w)$ for the length of a word
(thus $\ell(w) \in \NN_0$ for all $w \in {\cal A}$), we can define
(finite) subsets of ${\cal A}$ by
\begin{equation}
     {\cal A}_n \; := \; \{ w \in {\cal A} \;\, | \;\, \ell(w)=n \} \, .
\end{equation}
Here, ${\cal A}_0$ consists only of the empty word and
\begin{equation}
     {\cal A} \; = \; {\cal A}_{\infty} \; = \; 
     \bigcup_{n=0}^{\infty} {\cal A}_n
\end{equation}
which is sometimes also called the dictionary of the trivial language
(we have put no rules yet, so the above language simply consists of
all finite words in the alphabet $A$). We will generalize this
in a moment, but consider only situations where the number of words
of length $n$ behaves in such a way that its logarithm divided by $n$
defines a sequence that converges.

This is obviously so in the above case, where we have
\begin{equation}
       | {\cal A}_n |  \; = \;  x^n \, .
\end{equation}
Now, we can define the entropy $s = s(x)$ 
(it is actually an entropy density) through
\begin{equation}
      s(x)\; = \;\lim_{n\rightarrow\infty}\frac{\log(|{\cal A}_n|)}{n}\, .
\end{equation}
In our present case, we have of course
\begin{equation} 
      s(x) \; = \;  \log(x)
\end{equation}
which is a measure for the growth rate of $|{\cal A}_n|$ in $n$:
$|{\cal A}_n| = \exp(n\cdot s(x))$.

Now, let us introduce the concept of repeat-free or square-free
words. A word $w$ is called {\em square-free} if neither $w$ nor any
substring of it is a square, otherwise $w$ is said to contain a
square.  So, in an alphabet with two letters, $A=\{a,b\}$ say, $a$,
$b$, $ab$, $ba$, $aba$ and $bab$ are square-free while all other words
in these two letters are not. Consequently, as there are only finitely
many square-free words in two letters, the corresponding entropy is
zero. To make this more precise let us define
\begin{equation} \label{union1}
      {\cal A}_n \;  = \;  {\cal A}_n^+ \cup {\cal A}_n^-
\end{equation}
where ${\cal A}_n^-$ (${\cal A}_n^+$) denotes the subset of
all square-free words (square-containing words) of length $n$,
and the right hand side of (\ref{union1}) is clearly the
union of two disjoint sets. Consequently,
\begin{equation}
      | {\cal A}_n^+ | + | {\cal A}_n^- | \; = \;  |{\cal A}_n| \; = \; x^n
\end{equation}
and we introduce the abbreviation
\begin{equation}
       \omega_n^{\pm}  \; := \;  | {\cal A}_n^{\pm} |
\end{equation}
for convenience. There is always an implicit dependence on $x$,
the (finite) number of letters of the alphabet $A$, but we will often
suppress it when it is not needed.

One can now derive several properties of these numbers. It is clear
from Eq.~(\ref{union1}) that we have
\begin{equation} \label{sum1}
     \omega_n^+(x)  +  \omega_n^-(x)  \; = \;  x_{}^n
\end{equation}
and also the initial conditions
\begin{equation} \label{initial}
    \omega_0^-(x) = 1 \, , \;  \omega_1^-(x) = x \, , \;
    \omega_0^+(x) = \omega_1^+(x) = 0 \; .
\end{equation}
Observe also that $\omega_{2}^+(x) = x$, as there are precisely $x$
possibilities for words of type $aa$ etc.  If now, for $x$ fixed,
$\omega_n^-(x) = 0$ for some $n$, we must have $\omega_{n+m}^+(x) =
x^{n+m}$ for all $m \geq 0$. This is so because no continuation of a
square-containing word can become square-free and we can then use
Eq.~(\ref{sum1}).  This situation occurs with $x=2$, but not with any
larger $x$.  We can strengthen this type of argument to obtain, for
$n>0$,
\begin{equation} \label{greater1}
   \omega_{n+1}^+(x) \; \geq \;  x \cdot \omega_n^+(x)  +  \omega_n^-(x) \; .
\end{equation}
This is so because every word of ${\cal A}_n^+$ becomes, by adding
one arbitrary letter of the alphabet ($x$ possibilities), a word of 
${\cal A}_{n+1}^+$, while every word of ${\cal A}_n^-$ can be made
to one of ${\cal A}_{n+1}^+$ at least by repeating the last letter of it.

Now, with the initial conditions (\ref{initial}), 
repeated application of inequality (\ref{greater1}) 
shows ($n>1$)
\begin{equation} \label{greater2}
      \omega_n^+(x) \; \geq \;  x_{}^{n-1}  \; .
\end{equation}
It is possible to define the entropy $s_{}^+$ of the dictionary of
square-containing words, ${\cal A}_{\infty}^+$, which
exists as an ordinary limit. We obtain
\begin{prop} \label{prop1}
        The entropy of ${\cal A}_{\infty}^+$ equals that of 
        ${\cal A}_{\infty}^{}$: 
        $\quad s_{}^+(x) \, = \,  s(x) \, = \, \log(x)$.
\end{prop}
The proof is a direct application of inequality (\ref{greater2}):
\begin{displaymath}
    s_{}^+(x) \, = \, \lim_{n \rightarrow \infty}
       \frac{\log(\omega_n^+(x))}{n}  \, \geq \, 
       \log(x) \cdot \lim_{n \rightarrow \infty}
       \frac{n-1}{n} \, = \, \log(x) \, .
\end{displaymath}
On the other hand, we have $s_{}^+(x) \leq s(x) = \log(x)$,
from which the statement follows. $\Box$

For the other entropy, $s_{}^-(x)$, we have to prove existence
as an ordinary limit first. We do that in a slightly more general
form, following an argument given by Pleasants \cite{Pleasants}.
\begin{lemma} \label{limit}
   Let $a(n)$ be a sequence of positive integers with 
   $a(m+n) \leq a(m) \cdot a(n)$. Then the
   sequence defined through $h(n) := \log(a(n))/n$ is convergent.
\end{lemma}
{\sc Proof}:
Let us take two integers $N\geq n$ related by $N=qn+r$ with $0\leq r<n$.
We then have, with $x:=a(1)$,
\begin{equation}
   a(N) \; \leq \; \left(a(n)\right)^q \cdot a(r)
           \; \leq \; \left(a(n)\right)^q \cdot x^r
\end{equation}
and, consequently, 
\begin{equation}
   \frac{\log(a(N))}{N} \; \leq \;
    \frac{q \cdot \log(a(n))}{N} + \frac{r \cdot \log(x)}{N}
    \; < \;  \frac{\log(a(n))}{n} + \frac{\log(x)}{q} \, .
\end{equation}
This means $h(N) < h(n) + \mbox{\small $\frac{1}{q} \log(x)$}$
for any $N=q n+r$ with $0\leq r<n$. But then, we can fix $n$ and take 
the limit $N\rightarrow\infty$ which also implies $q\rightarrow\infty$.
This gives $\limsup_{N\rightarrow\infty} h(N) \leq h(n)$.
The last equation is valid for all $n \in \NN$, so we also have
\begin{equation}
      \limsup_{N\rightarrow\infty} h(N)
     \; \leq \; \liminf_{n\rightarrow\infty} h(n)
\end{equation}
which means that both must be equal and the limit exists. $\Box$

If $a(n)$ is only a sequence of non-negative integers, but still
with $a(m+n) \leq a(m) a(n)$, one either has $a(n)>0$ for all $n$
or, if $a(n_0^{})=0$ for some $n_0^{}$, one has $a(n)=0$ also for
all $n>n_0^{}$ due to submultiplicativity.
In the latter case, we follow the usual convention
and define $h(n)=0$ which results in 
$\lim_{n\rightarrow\infty} h(n) = 0$, i.e.\ vanishing entropy.

It is clear that the above Lemma is not the most general formulation of the
statement, but it is sufficient for our needs. The language of
square-free words is {\em subword-closed}, i.e., no new substring
of length $n$ can occur in any word of length $>n$ \cite{Pleasants}. 
Consequently, our numbers $\omega_n^-(x)$ are such that the lemma applies.
We have thus shown
\begin{prop}
  The entropy $s^-(x)$ of ${\cal A}^-_{\infty}(x)$ exists as a limit.
\end{prop}

We so far only know the trivial inequality
\begin{equation} \label{entropy1}
      0 \; \leq \; s_{}^-(x)  \; \leq \;  \log(x)
\end{equation}
with $s_{}^-(x) = 0$ for $x=1$ and $x=2$. Since we have at most
$x-1$ possibilities to make a square-free word of length $n$
into one of length $n+1$ by adding a letter, we also have ($n \geq 1$)
\begin{equation} \label{greater3}
      \omega_{n+1}^-(x)  \; \leq \;  (x-1) \cdot \omega_n^-(x)
\end{equation}
from which we can improve the upper bound of (\ref{entropy1}) to
\begin{equation} \label{entropy2}
    s_{}^-(x)  \; \leq \;  \log(x-1) \; .
\end{equation}

Of course, we can further improve the upper bound by considering
the possibilities of appending (square-free) words which consist
of more than one letter. In practice, this amounts to actually
counting the number of square-free words of a certain length.
One obtains
\begin{equation} \label{greater3a}
\omega_{n+k}^-(x) \; \leq \; \frac{\omega_{k+j}^-(x)}{\omega_{j}^-(x)}
\omega_n^-(x)
\end{equation}
where $j\in\{0,1,2\}$ and $n\geq j$ and $k\geq0$ are arbitrary.
This expression follows by extending square-free words of length $n$
with an overlapping square-free word of length $k+j$.
By restricting the length of the overlap to $j\leq 2$
all possibilities appear with the same frequency
(by symmetry). For the entropy, this yields upper bounds
\begin{equation} \label{entropy2a}
s_{}^-(x)  \; \leq \;  \frac{1}{k} \left( \log(\omega_{k+j}^-(x)) -
\log(\omega_{j}^-(x))\right)
\end{equation}
which clearly gives the strongest bound for $j=2$, with
$\omega_{2}^-(x)=x(x-1)$.

Now, a more interesting point is the question for a lower bound
of the entropy $s_{}^-(x)$. We know it is zero for $x=1$ and $x=2$.
As we will show, it is strictly positive for other $x$, i.e.~for $x>2$.
The case $x=3$ will play a special role, but let us first give some
simple results. Here, we rely on the well-known fact that there
exists at least one square-free word of infinite length in three letters,
compare Ref.~\cite{Lothaire} and references therein. But then, we can 
directly show
\begin{theorem} 
     For $x \geq 3$, the entropy of ${\cal A}_{\infty}^-$ has 
     a lower bound: $\quad s_{}^-(x) \, \geq \, \frac{1}{3} \log(x-2)$.
 
     \noindent In particular, the entropy is strictly positive for $x>3$.
\end{theorem}
{\sc Proof}: Let $w$ be a square-free word of infinite length in 3 letters,
$\{a,b,c\}$ say, which we know to exist from Refs.~\cite{Thue1,Lothaire}. 
Let $w^{}_n$ be the subword made from the first $n$ letters of $w$
which is also square-free and contains $m_a,m_b,m_c$ letters of 
type $a,b,c$, respectively, where $m_a + m_b + m_c = n$. 
Now, let $\{d_1, d_2, \ldots, d_p\}$ be $p$ new letters, $p > 0$. 
If we fix all $b$'s and $c$'s in $w^{}_n$, we have $(p+1)^{m_a}$
possibilities to make square-free words of length $n$ in four
letters by successive replacement of any $a$ in $w$ by an
element of $\{a,d_1,  \ldots ,d_p\}$. 
We can analogously proceed for the other two letters, $b$ and $c$,
through fixing $a,c$ and $a,b$, respectively. 
In this setup, we have $x=p+3$, and we can conclude (for $x\geq3$)
\begin{equation} \label{greater4}
     \omega_n^-(x) \; \geq \;  (x-2)^{m_a} + (x-2)^{m_b} + (x-2)^{m_c}
      \;  \geq  \;  3 \cdot (x-2)_{}^{n/3}
\end{equation}
where the second inequality is a standard result from calculus.
But from this, we immediately get the inequality
\begin{equation} \label{bound1}
      s_{}^-(x)  \;  \geq  \;  \frac{1}{3} \log(x-2) + 
              \lim_{n \rightarrow \infty}  \frac{3}{n}  \; = \;
              \frac{1}{3} \log(x-2)
\end{equation}
from which the statement follows. $\Box$

We cannot gain anything about $\varepsilon := s_{}^-(3)$ this way,
although, as we will see, precisely this $\varepsilon$ is important.
Nevertheless, we can do better than (\ref{bound1}). In the above
argument, we started from an infinite word $w$ in three letters.
Instead, we can also apply the same type of argument for the
step from $x$ to $x+1$ letters: fixing $x-1$ letters, we stay with
two possibilities for the replacement of every occurrence of the 
remaining letter, and this can be done in $x$ different ways.
As it applies essentially to every word separately,
the number of possibilities behaves almost multiplicatively, 
i.e.\ it grows like
\begin{equation}
       \omega_n^-(x+1) \; \sim \;   x \cdot 2^{n/x}  \cdot \omega_n^-(x) \, .
\end{equation}
We cannot write $\geq$ instead of $\sim$ here, 
as one does in fact multiply count several words.
 
To avoid this, we have to discard all words of ${\cal A}_n^-(x)$
that do not contain all letters. Let us denote the number of square-free
words of length $n$ in {\em exactly} $x$ letters by $\psi_n(x)$.
Clearly, $\psi_n(0) = \delta_{n,0}$, $\psi_n(1) = \delta_{n,1}$
and $\psi_n(2) = 2(\delta_{n,2}+\delta_{n,3})$. 
Also, $\psi_n^{} (3) = \omega_n^- (3) > 0$ for all $n>3$.
Furthermore, we have
\begin{equation} \label{psi1}
     \omega_n^-(x) \; = \; \sum_{k=0}^x
       \binco{x^{}}{k} \psi_n^{}(k)
\end{equation}
and also, obviously, $\omega_n^-(x) \geq \psi_n^{}(x)$ for all $n,x$.
For $x>n$, one has $\psi_{n}(x) = 0$, and $\psi_{x}(x) = x!$, the number
of permutations of $x$ symbols. By simple counting, one finds
$\psi_{x+1}(x) = x! x(x-1)/2$.
A little less obvious is the inversion of Eq.~(\ref{psi1}) in the form
\begin{equation} \label{psi2}    
      \psi_n^{}(x) \; = \; \sum_{k=0}^x  (-1)^{x\!-\!k}
       \binco{x^{}}{k} \omega_n^-(k) \, .
\end{equation}
The number $\psi_n(x)$ is a multiple of $x!$ since any permutation
of the $x$ letters transforms a square-free word in exactly $x$
letters into another one ot this type.
The advantage of the new numbers is that we can go from $\psi_n(x)$
to $\psi_n(x+1)$ without any double counting, i.e., two different
square-free words of length $n$ with all $x$ letters in them 
give automatically two disjoint sets of square-free words of 
length $n$ in $x+1$ letters by the above procedure. 
If we observe that we must introduce at least one new letter,
we see that, for $x\geq 1$ and $n\geq x+1$, we get the inequality
\begin{equation} \label{psi3}
       \psi_n (x+1) \; \geq \;    \frac{x}{2} 
          \cdot 2^{n/x}   \cdot \psi_n (x) \, .
\end{equation}
This is helpful because we have:
\begin{lemma}
Let $x\geq 3$ be fixed. Then,
the sequences $\psi^{}_n (x)$ and $\omega^-_n (x)$ 
have the same exponential growth in $n$:
\begin{displaymath} 
    \lim_{n \rightarrow \infty}
       \frac{\log(\psi^{}_n (x))}{n}   \; = \;
    \lim_{n \rightarrow \infty}
       \frac{\log(\omega^-_n (x))}{n}  \, .
\end{displaymath}
\end{lemma}
{\sc Proof}: Iterating Eq.~(\ref{psi3}) one obtains 
(for $0 < k < x$ and $n\geq x$) the inequality
\begin{equation} \label{psi4}
   \psi_n (k) \; \leq \;  \psi_n (x) \cdot \left(
    \frac{2^{x-k}_{}}{2_{}^{n ({1\over k} + \cdots + {1\over x-1})} }
    \cdot \frac{(k-1)!}{(x-1)!} \right) \, .
\end{equation}
Consequently, we obtain, from Eq.~(\ref{psi1}), the inequality
\begin{equation}
   \omega_n^- (x)  \; \leq \;  \psi_n^{} (x) \left[ 1 +
     \sum_{k=1}^{x-1} \frac{x}{k} \cdot \frac{2^{x-k}_{}}{(x-k)!}
     \cdot \frac{1}{2_{}^{n ({1\over k} + \cdots + {1\over x-1}) }}
     \right] \, .
\end{equation}
Since every single term under the sum is certainly not bigger than $2x$,
we finally get
\begin{equation} \label{psi5}
   \psi_n^{} (x) \; \leq \; \omega_n^- (x) \; \leq \;
   (1 + 2 x(x\!-\!1)) \cdot \psi_n^{} (x)  \; \leq \;
   (1 + 2 x(x\!-\!1)) \cdot \omega_n^- (x)
\end{equation}
from which the statement easily follows. $\Box$

{}From this Lemma and from Eq.~(\ref{psi3}) we see 
that the entropic contributions are additive:
\begin{equation} \label{entropy3}
       s_{}^-(x+1) \;  \geq  \;     s_{}^-(x) + \frac{\log(2)}{x}
\end{equation}
which is valid for $x\geq 3$.
In fact, repeating the argument of (\ref{entropy3}), 
we immediately arrive at
\begin{theorem} \label{thm}
Let $x>3$. The entropy of the dictionary ${\cal A}_{\infty}^{-}(x)$ fulfills:
\begin{displaymath}
       \varepsilon + \log(2) \cdot \left( \frac{1}{3} + \frac{1}{4} + \cdots + 
       \frac{1}{x-1} \right)   \; \leq \;
      s_{}^-(x) \; \leq \;  \log(x-1) \, .
\end{displaymath} \end{theorem}
Here, and in what follows, we will always use $\varepsilon$ for $s_{}^-(3)$.
Of course, one can now use the formula
\begin{equation} \label{euler}
       \sigma_m \; := \;
      1 + \frac{1}{2} + \frac{1}{3} + \cdots + \frac{1}{m} - \log(m)
      \; \stackrel{m \rightarrow \infty}{\longrightarrow} \;
       \gamma \simeq 0.5771\ldots
\end{equation}
for Euler's constant and the fact that the sequence $(\sigma_m)_{m \in \NNs}$
is strictly decreasing to simplify the inequality of Thm.~(\ref{thm}) for
larger values of $x$, while for small values it is better to stick to 
the finite sum.

Again, we have seen that $\varepsilon$ plays a special role. In fact, although
this does not follow from any simple inequality of the above type,
$\varepsilon$ is strictly positive \cite{Elser,Brandenburg,Brinkhuis}.
Brandenburg \cite{Brandenburg} shows that
\begin{equation}
\omega_{22n}^-(3) \; \geq \; 2^n \omega_n^-(3)
\end{equation}
and concludes from this that
\begin{equation}
\omega_n^-(3) \; \geq \; 6 \cdot 2^{n/22} \, .
\end{equation}
In fact, using the {\em existence} of the limit (which we have shown
above), we can improve the argument slightly
(cf.\ Ref.~\cite{Elser,Brinkhuis}) to obtain a strict lower bound
for the entropy of square-free words on three letters:
\begin{theorem}
The entropy $\varepsilon = s^-(3)$ is strictly positive:
\begin{equation}
\varepsilon \; \geq \; \frac{1}{21}\log(2) \; \simeq \; 0.033007\ldots
\end{equation}
\end{theorem}

Since this kind of result has been described several times already
\cite{Elser,Brandenburg,Brinkhuis}, we shall not repeat the proof.
The idea behind it is the following: one tries to find a set
of substitution rules which map square-free words into square-free
words of increased length and simultaneously allow some free choice
to do so in each step -- which accounts for the exponential growth.
It is not clear whether one can significantly improve the lower
bound in this way so as to approach our numerical estimate
of 
\begin{equation}
   \varepsilon \; \simeq \; 0.263719 \,(1) \; .
\end{equation}

\section{Some results for three letters}
\setcounter{equation}{0}

Let us describe the case $x=3$ in more detail.
We can write the set of square-free words in three letters
as a disjoint union of three subsets
\begin{equation}
{\cal A}_n^- \; = \; {\cal A}_n^{(0)} \cup {\cal A}_n^{(1)} 
                     \cup {\cal A}_n^{(2)} \; .
\end{equation}
Here, ${\cal A}_n^{(0)}$ denotes the set of what we call {\em stop-words}. 
These are characterized by the property that by appending {\em any} letter 
of the alphabet $A$ one obtains a square-containing word. In the same spirit,
${\cal A}_n^{(1)}$ and ${\cal A}_n^{(2)}$ are defined as the sets of 
square-free words which allow, respectively, one and two extensions to a 
square-free word of length $n+1$. Introducing the notation
\begin{equation}
\omega_n^{(k)} \; := \;  | {\cal A}_n^{(k)} | \; , \quad k\in\{0,1,2\} \; ,
\end{equation}
this implies the relation
\begin{equation}
\omega_{n+1}^- \; = \; \sum_{k=0}^{2} \omega_{n+1}^{(k)}
               \; = \; \sum_{k=0}^{2} k\cdot\omega_n^{(k)}
               \; = \; \omega_n^{(1)} + 2\cdot\omega_n^{(2)} \; .
\end{equation}
Hence the growth rate is given by
\begin{equation} 
\frac{\omega_{n+1}^-}{\omega_{n}^-} \; = \;
1 + \frac{\omega_n^{(2)}-\omega_n^{(0)}}{\omega_{n}^-} \; .
\end{equation}
Since the left-hand side converges to a finite value 
in the limit $n\rightarrow\infty$, so does the right-hand side, and
by the positivity of the entropy we see that convergence of the
sequence $\omega_n^{(0)}/\omega_{n}^-$ would imply the convergence
of the ratio $\omega_n^{(2)}/\omega_n^-$ to a finite non-zero limit.

In order to gain some insight into the actual behaviour of these
quantities we used the computer to investigate square-free words in three
letters for lengths up to 90. The results are presented in Tables~1 and 2. 

In Table~1, we list the number of square-free words $\omega_n^-$,
approximants to the entropy obtained from the ratio of successive
values, and upper limits to the entropy using Eq.~(\ref{entropy2a}).
{}From these, we extract 
\begin{equation}
     \varepsilon\simeq 0.263719\,(1)
\end{equation}
as the approximate value of the entropy of square-free words on three letters,
where the figure in parentheses denotes the estimated uncertainty in the
last digit.

It is striking that the logarithm of the ratio obviously approaches
the limit value much faster than the value in the last column.
This in fact suggests that the asymptotic behaviour looks as follows
\begin{equation}
  \log(\omega_n^-) \; \sim \; \varepsilon \cdot n + a + o(n^{-1}) 
  \quad (n \rightarrow \infty)
\end{equation}
where the constant term can be estimated as $a \simeq 2.5438965$.
It is interesting that the next order seems to fall so quickly, 
which also indicates that no logarithmic corrections are present.

Table~2 contains the values of $\omega_n^{(k)}$ ($k=0,1,2$) and their ratios
with $\omega_n^-$. Apparently, the ratios converge as $n\rightarrow\infty$,
which means that all three subsets have in fact the same entropy as the
set of square-free words itself. For the ratios, we estimate
\begin{equation}
\frac{\omega_n^{(0)}}{\omega_n^-} \;\leadsto\; 0.036837 \, ,
\qquad
\frac{\omega_n^{(1)}}{\omega_n^-} \;\leadsto\; 0.624564 \, ,
\qquad
\frac{\omega_n^{(2)}}{\omega_n^-} \;\leadsto\; 0.338599 \, ,
\end{equation}
in the limit $n\rightarrow\infty$, where the uncertainty is about one
figure in the last digit.

We note that the stop-words still show an interesting structure if
one looks at the three lengths of squares that one obtains on
appending the three different letters. Apparently, stop-words with
certain fixed sets of periods still occur with the same entropy
density, whereas other periods are limited to
``symmetric'' stop-words that cannot be extended in any
direction and therefore show up in finitely many stop-words only.
As an example for this, we mention the shortest stop-words
($abacaba$ and those obtained from permutation of letters,
where $a$, $b$ and $c$ denote the three letters) which
result in squares of lengths 1, 2 and 4. Here, it is easy to see that
these are the only words with this property.

\clearpage

{\footnotesize
\centerline{Table 1: Number of square-free words in three letters; 
estimates and upper limits for their entropy.}
\begin{center}
\addtolength{\tabcolsep}{2mm}
\begin{tabular}{|r|r|c|c|}
\hline
\multicolumn{1}{|c}{\rud $n$} &
\multicolumn{1}{|c}{$\omega_n^-$} &
\multicolumn{1}{|c}{$\log(\omega_n^-/\omega_{n-1}^-)$} &
\multicolumn{1}{|c|}{$\frac{\log(\omega_n^-/6)}{n-2}$} \\
\hline\ru
1  &            3 &            &            \\
2  &            6 & 0.69314718 &            \\
3  &           12 & 0.69314718 & 0.69314718 \\
4  &           18 & 0.40546511 & 0.54930615 \\
5  &           30 & 0.51082562 & 0.53647929 \\
6  &           42 & 0.33647224 & 0.48647752 \\
7  &           60 & 0.35667494 & 0.46051702 \\
8  &           78 & 0.26236426 & 0.42749155 \\
9  &          108 & 0.32542240 & 0.41291025 \\
10 &          144 & 0.28768207 & 0.39725673 \\
11 &          204 & 0.34830669 & 0.39181784 \\
12 &          264 & 0.25782911 & 0.37841898 \\
13 &          342 & 0.25886163 & 0.36755010 \\
14 &          456 & 0.28768207 & 0.36089444 \\
15 &          618 & 0.30399565 & 0.35651761 \\
16 &          798 & 0.25562014 & 0.34931064 \\
17 &         1044 & 0.26870617 & 0.34393701 \\
18 &         1392 & 0.28768207 & 0.34042108 \\
19 &         1830 & 0.27357440 & 0.33648893 \\
20 &         2388 & 0.26614023 & 0.33258069 \\
21 &         3180 & 0.28642500 & 0.33015144 \\
22 &         4146 & 0.26526282 & 0.32690698 \\
23 &         5418 & 0.26758273 & 0.32408205 \\
24 &         7032 & 0.26074442 & 0.32120302 \\
25 &         9198 & 0.26851491 & 0.31891227 \\
26 &        11892 & 0.25687984 & 0.31632757 \\
27 &        15486 & 0.26407048 & 0.31423730 \\
28 &        20220 & 0.26673582 & 0.31241032 \\
29 &        26424 & 0.26760047 & 0.31075069 \\
30 &        34422 & 0.26442321 & 0.30909613 \\
31 &        44862 & 0.26489522 & 0.30757198 \\
32 &        58446 & 0.26451214 & 0.30613664 \\
33 &        76122 & 0.26423407 & 0.30478492 \\
34 &        99276 & 0.26556653 & 0.30355936 \\
35 &       129516 & 0.26590058 & 0.30241820 \\
36 &       168546 & 0.26340428 & 0.30127072 \\
37 &       219516 & 0.26421641 & 0.30021203 \\
38 &       285750 & 0.26369218 & 0.29919758 \\
39 &       372204 & 0.26432479 & 0.29825506 \\
40 &       484446 & 0.26356388 & 0.29734215 \\
41 &       630666 & 0.26377043 & 0.29648134 \\
42 &       821154 & 0.26393426 & 0.29566768 \\
43 &      1069512 & 0.26424708 & 0.29490131 \\
44 &      1392270 & 0.26373304 & 0.29415920 \\ \rd
45 &      1812876 & 0.26397903 & 0.29345734 \\
\hline               
\end{tabular}
\end{center}
\addtolength{\tabcolsep}{-2mm}
}
\clearpage

{\footnotesize
\centerline{Table 1: (continued)}
\addtolength{\tabcolsep}{2mm}
\begin{center}
\begin{tabular}{|r|r|c|c|}
\hline
\multicolumn{1}{|c}{\rud $n$} &
\multicolumn{1}{|c}{$\omega_n^-$} &
\multicolumn{1}{|c}{$\log(\omega_n^-/\omega_{n-1}^-)$} &
\multicolumn{1}{|c|}{$\frac{\log(\omega_n^-/6)}{n-2}$} \\
\hline\ru           
46 &      2359710 & 0.26362420 & 0.29277931 \\
47 &      3072486 & 0.26394828 & 0.29213862 \\
48 &      4000002 & 0.26380786 & 0.29152274 \\
49 &      5207706 & 0.26384459 & 0.29093384 \\
50 &      6778926 & 0.26367923 & 0.29036604 \\
51 &      8824956 & 0.26376493 & 0.28982316 \\
52 &     11488392 & 0.26375352 & 0.28930176 \\
53 &     14956584 & 0.26381447 & 0.28880201 \\
54 &     19470384 & 0.26374294 & 0.28832011 \\
55 &     25346550 & 0.26374808 & 0.28785649 \\
56 &     32996442 & 0.26375711 & 0.28741020 \\
57 &     42957300 & 0.26380686 & 0.28698105 \\
58 &     55921896 & 0.26374940 & 0.28656620 \\
59 &     72798942 & 0.26374542 & 0.28616583 \\
60 &     94766136 & 0.26371071 & 0.28577868 \\
61 &    123368406 & 0.26376292 & 0.28540553 \\
62 &    160596120 & 0.26371759 & 0.28504406 \\
63 &    209059806 & 0.26372772 & 0.28469461 \\
64 &    272143380 & 0.26370870 & 0.28435613 \\
65 &    354271314 & 0.26373398 & 0.28402880 \\
66 &    461181036 & 0.26372763 & 0.28371159 \\
67 &    600356406 & 0.26373282 & 0.28340422 \\
68 &    781520994 & 0.26371852 & 0.28310596 \\
69 &   1017362166 & 0.26372643 & 0.28281671 \\
70 &   1324371090 & 0.26372453 & 0.28253594 \\
71 &   1724034504 & 0.26372949 & 0.28226338 \\       
72 &   2244278358 & 0.26371684 & 0.28199843 \\
73 &   2921521164 & 0.26372040 & 0.28174100 \\
74 &   3803130042 & 0.26372000 & 0.28149071 \\
75 &   4950798954 & 0.26372455 & 0.28124733 \\
76 &   6444761514 & 0.26371866 & 0.28101046 \\
77 &   8389549680 & 0.26371921 & 0.28077991 \\
78 &  10921197582 & 0.26371879 & 0.28055542 \\
79 &  14216853012 & 0.26372246 & 0.28033681 \\
80 &  18506985300 & 0.26372015 & 0.28012378 \\
81 &  24091726728 & 0.26372025 & 0.27991614 \\
82 &  31361678988 & 0.26371824 & 0.27971366 \\
83 &  40825520274 & 0.26372065 & 0.27951622 \\
84 &  53145145482 & 0.26371938 & 0.27932357 \\
85 &  69182396616 & 0.26371968 & 0.27913558 \\
86 &  90058945560 & 0.26371796 & 0.27895203 \\
87 & 117235364616 & 0.26371917 & 0.27877282 \\
88 & 152612592438 & 0.26371906 & 0.27859778 \\
89 & 198665414208 & 0.26371944 & 0.27842676 \\ \rd
90 & 258615015792 & 0.26371846 & 0.27825962 \\
\hline                                        
\end{tabular}
\end{center}
\addtolength{\tabcolsep}{-2mm}
}
\clearpage

{\footnotesize
\centerline{Table 2: Number and relative frequence of square-free words 
 in three letters that allow 0, 1, and 2 extensions, respectively.}
\addtolength{\tabcolsep}{1mm}
\begin{center}
\begin{tabular}{|r|r|r|r|c|c|c|}
\hline
\multicolumn{1}{|c}{\rud $n$} &
\multicolumn{1}{|c}{$\omega_n^{(0)}$} &
\multicolumn{1}{|c}{$\omega_n^{(1)}$} &
\multicolumn{1}{|c}{$\omega_n^{(2)}$} &
\multicolumn{1}{|c}{$\omega_n^{(0)}/\omega_n^-$} &
\multicolumn{1}{|c}{$\omega_n^{(1)}/\omega_n^-$} &
\multicolumn{1}{|c|}{$\omega_n^{(2)}/\omega_n^-$} \\
\hline\ru                                    
1 &         0&           0&          3&0.00000000&0.00000000&1.00000000\\
2 &         0&           0&          6&0.00000000&0.00000000&1.00000000\\
3 &         0&           6&          6&0.00000000&0.50000000&0.50000000\\
4 &         0&           6&         12&0.00000000&0.33333333&0.66666667\\
5 &         0&          18&         12&0.00000000&0.60000000&0.40000000\\
6 &         0&          24&         18&0.00000000&0.57142857&0.42857143\\
7 &         6&          30&         24&0.10000000&0.50000000&0.40000000\\
8 &         0&          48&         30&0.00000000&0.61538462&0.38461538\\
9 &         0&          72&         36&0.00000000&0.66666667&0.33333333\\
10&         0&          84&         60&0.00000000&0.58333333&0.41666667\\
11&         6&         132&         66&0.02941176&0.64705882&0.32352941\\
12&         6&         174&         84&0.02272727&0.65909091&0.31818182\\
13&         6&         216&        120&0.01754386&0.63157895&0.35087719\\
14&         6&         282&        168&0.01315789&0.61842105&0.36842105\\
15&        24&         390&        204&0.03883495&0.63106796&0.33009709\\
16&        24&         504&        270&0.03007519&0.63157895&0.33834586\\
17&        24&         648&        372&0.02298851&0.62068966&0.35632184\\
18&        36&         882&        474&0.02586207&0.63362069&0.34051724\\
19&        54&        1164&        612&0.02950820&0.63606557&0.33442623\\
20&        54&        1488&        846&0.02261307&0.62311558&0.35427136\\
21&       120&        1974&       1086&0.03773585&0.62075472&0.34150943\\
22&       138&        2598&       1410&0.03328509&0.62662808&0.34008683\\      
23&       216&        3372&       1830&0.03986711&0.62236988&0.33776301\\
24&       240&        4386&       2406&0.03412969&0.62372014&0.34215017\\
25&       384&        5736&       3078&0.04174821&0.62361383&0.33463796\\
26&       444&        7410&       4038&0.03733602&0.62310797&0.33955600\\
27&       528&        9696&       5262&0.03409531&0.62611391&0.33979078\\
28&       690&       12636&       6894&0.03412463&0.62492582&0.34094955\\
29&       966&       16494&       8964&0.03655767&0.62420527&0.33923706\\
30&      1236&       21510&      11676&0.03590727&0.62489106&0.33920167\\
31&      1602&       28074&      15186&0.03570951&0.62578574&0.33850475\\
32&      2112&       36546&      19788&0.03613592&0.62529514&0.33856894\\
33&      2712&       47544&      25866&0.03562702&0.62457634&0.33979664\\
34&      3522&       61992&      33762&0.03547685&0.62444095&0.34008220\\
35&      4818&       80850&      43848&0.03720004&0.62424720&0.33855277\\
36&      6150&      105276&      57120&0.03648856&0.62461287&0.33889858\\
37&      8094&      137094&      74328&0.03687203&0.62452851&0.33859946\\
38&     10452&      178392&      96906&0.03657743&0.62429396&0.33912861\\
39&     13854&      232254&     126096&0.03722152&0.62399652&0.33878196\\
40&     17784&      302658&     164004&0.03670997&0.62475075&0.33853928\\
41&     23082&      394014&     213570&0.03659940&0.62475859&0.33864201\\
42&     29970&      512856&     278328&0.03649742&0.62455520&0.33894738\\
43&     39438&      667878&     362196&0.03687476&0.62446985&0.33865539\\
44&     51030&      869604&     471636&0.03665237&0.62459437&0.33875326\\ \rd
45&     66792&     1132458&     613626&0.03684312&0.62467483&0.33848206\\
\hline                  
\end{tabular}
\end{center}
\addtolength{\tabcolsep}{-1mm}
}
\clearpage

{\footnotesize
\centerline{Table 2: (continued)}
\addtolength{\tabcolsep}{1mm}
\begin{center}
\begin{tabular}{|r|r|r|r|c|c|c|}
\hline
\multicolumn{1}{|c}{\rud $n$} &
\multicolumn{1}{|c}{$\omega_n^{(0)}$} &
\multicolumn{1}{|c}{$\omega_n^{(1)}$} &
\multicolumn{1}{|c}{$\omega_n^{(2)}$} &
\multicolumn{1}{|c}{$\omega_n^{(0)}/\omega_n^-$} &
\multicolumn{1}{|c}{$\omega_n^{(1)}/\omega_n^-$} &
\multicolumn{1}{|c|}{$\omega_n^{(2)}/\omega_n^-$} \\
\hline\ru                                        
46&     86502&     1473930&     799278&0.03665789&0.62462336&0.33871874\\
47&    113064&     1918842&    1040580&0.03679887&0.62452425&0.33867689\\
48&    147036&     2498226&    1354740&0.03675898&0.62455619&0.33868483\\
49&    191952&     3252582&    1763172&0.03685922&0.62457097&0.33856980\\
50&    249390&     4234116&    2295420&0.03678901&0.62459983&0.33861116\\
51&    324852&     5511816&    2988288&0.03681061&0.62457150&0.33861789\\
52&    422712&     7174776&    3890904&0.03679471&0.62452395&0.33868134\\
53&    550758&     9341268&    5064558&0.03682378&0.62455892&0.33861729\\
54&    716454&    12161310&    6592620&0.03679712&0.62460555&0.33859733\\
55&    932592&    15831474&    8582484&0.03679365&0.62460074&0.33860561\\
56&   1213602&    20608380&   11174460&0.03677978&0.62456370&0.33865651\\
57&   1582026&    26828652&   14546622&0.03682787&0.62454232&0.33862980\\
58&   2058528&    34927794&   18935574&0.03681077&0.62458172&0.33860751\\
59&   2681796&    45468156&   24648990&0.03683839&0.62457166&0.33858995\\
60&   3488478&    59186910&   32090748&0.03681144&0.62455760&0.33863096\\
61&   4545588&    77049516&   41773302&0.03684564&0.62454820&0.33860616\\
62&   5914926&   100302582&   54378612&0.03683106&0.62456417&0.33860477\\
63&   7701792&   130572648&   70785366&0.03684014&0.62457079&0.33858907\\
64&  10021482&   169972482&   92149416&0.03682427&0.62456960&0.33860613\\
65&  13049082&   221263428&  119958804&0.03683358&0.62455926&0.33860716\\
66&  16985274&   288035118&  156160644&0.03682995&0.62455976&0.33861029\\
67&  22114344&   374963130&  203278932&0.03683536&0.62456755&0.33859709\\
68&  28782414&   488114994&  264623586&0.03682872&0.62457055&0.33860074\\
69&  37472418&   635408406&  344481342&0.03683292&0.62456461&0.33860247\\
70&  48778746&   827150184&  448442160&0.03683163&0.62456074&0.33860763\\
71&  63510756&  1076769138&  583754610&0.03683845&0.62456357&0.33859799\\  
72&  82666266&  1401703020&  759909072&0.03683423&0.62456736&0.33859840\\
73& 107616300&  1824679686&  989225178&0.03683571&0.62456494&0.33859935\\
74& 140084994&  2375291142& 1287753906&0.03683413&0.62456217&0.33860370\\
75& 182377848&  3092080698& 1676340408&0.03683806&0.62456196&0.33859998\\
76& 237398214&  4025176920& 2182186380&0.03683584&0.62456569&0.33859847\\
77& 309038124&  5239825530& 2840686026&0.03683608&0.62456577&0.33859815\\
78& 402276216&  6820989720& 3697931646&0.03683444&0.62456426&0.33860130\\
79& 523700664&  8879319396& 4813832952&0.03683661&0.62456293&0.33860046\\
80& 681718896& 11558806080& 6266460324&0.03683576&0.62456450&0.33859974\\
81& 887460042& 15046854384& 8157412302&0.03683671&0.62456521&0.33859808\\
82&1155219294& 19587399114&10619060580&0.03683538&0.62456475&0.33859987\\
83&1503883698& 25498127670&13823508906&0.03683685&0.62456345&0.33859970\\
84&1957680408& 33192533532&17994931542&0.03683649&0.62456379&0.33859972\\
85&2548490760& 43208866152&23425039704&0.03683727&0.62456446&0.33859827\\
86&3317442636& 56247641232&30493861692&0.03683635&0.62456473&0.33859892\\
87&4318568760& 73220999274&39695796582&0.03683674&0.62456409&0.33859917\\
88&5621734092& 95316302484&51674555862&0.03683663&0.62456381&0.33859955\\ \rd
89&7318253526&124079305572&67267855110&0.03683708&0.62456420&0.33859872\\ 
\hline                                        
\end{tabular}
\end{center}
\addtolength{\tabcolsep}{-1mm}
}
\clearpage

\section{Basics of an algebraic approach}
\setcounter{equation}{0}

It is now time to attack the square-free words with some algebra.
To do so, we prefer to change the notation 
\begin{equation} \label{poly1}
     P_n(x)  \; := \;  | {\cal A}_n^- (x) |  \; = \;  \omega_n^- (x)
\end{equation}
because we will not talk about square-containing words any more.
Clearly, $P_n(x)$ is always an integer, and 
we also know that, for $n \in \NN$, $P_n(x) \leq x^n$, with equality
only for $n=0$ and $n=1$.
It is straight-forward to calculate the first cases explicitly
\begin{eqnarray} \label{poly2}
      P_0(x) \, = \, 1 \, , \;
      P_1(x) \, = \, x \, , \;
      P_2(x) \, = \, x (x\!-\!1)  \, , \; & & \nonumber \\
      P_3(x) \, = \, x (x\!-\!1)^2 \, , \;
      P_4(x) \, = \, x^2 (x\!-\!1)  (x\!-\!2) \, .   & &
\end{eqnarray}
The $P_n(x)$ are polynomials in $x$, which can be shown by induction.
If we go to $P_{n+1}(x)$ we can recursively build it from $(x-1)P_n(x)$
corrected by lower order terms that are sums of products of $P_k(x)$
with $k<n$ -- hence we stay with a polynomial.

Let us consider these polynomials in more detail.
If $x > 2$ and $n \geq 2$, we can fix the first two letters when we want
to count the square-free letters of length $n$ as $ab$ say. For
this we have $x (x\!-\!1)$ possibilities, each of which must have
equally many square-free continuations.
Similarly, let $x>3$ and $n\geq 4$. The start can be of the form
$abc$ (where one has $x (x\!-\!1) (x\!-\!2)$ possibilities, each with
equally many square-free continuations) or of the form $abac$
(again $x (x\!-\!1) (x\!-\!2)$ possibilities). The set of words obtained from
these two classes of possibilities is disjoint, so we can conclude
\begin{prop}
  Let $P_n(x)$ be the number of square-free words of length
  $n$ in an alphabet of $x$ letters. Then, for $n>1$, 
  $P_n(x)$ is a multiple of $x (x\!-\!1)$, while for $n>3$,
  it is a multiple of $x (x\!-\!1) (x\!-\!2)$.
  Furthermore, $P_n(x)$ is a polynomial in $x$ of order $n$,
  with integer coefficients and leading coefficient $1$.
\end{prop}
There is an alternative way to see that the last statement of the lemma 
is correct. 
Clearly, the number of all words of length $n$ in $x$ letters is just $x^n$, 
and from this we have to subtract the number of words which contain 
at least one square. 
Necessarily, demanding that a word contains a square of a certain length
(and no square of shorter length) means that one fixes a number of letters 
to coincide (and certain others to be different), thus all
the corresponding terms are of lower order in $x$. Also, they all have
integer coefficients since these are nothing but combinatorial factors.
Furthermore, as $P_n(x) \leq x^n$, the coefficient of the leading term
of $P_n(x)-x^n$ has to be negative.

So far, we did not manage to find the generating function for the polynomials
$P_n(x)$ -- and there are the usual indications that this might be a 
very difficult task: quite probably it will be one of those functions 
that are not analytically continuable beyond its circle of convergence. 
We assume this due to various unsuccessful attempts to find
generating functions for the numbers $\omega_n^-(x)$
by means of standard algebraic program packages -- although
it is certainly not conclusive.
So, we have tried to find a reasonable
approximation scheme. Here, we observe, from the explicit
counting and the determination of the $P_n(x)$ up to $n=15$, 
that the recurrence of the polynomials, for $n>2$, look as follows
\begin{equation} \label{recpoly}
    P_{n+1}(x) \; = \; (x-1) \cdot P_n(x) - P_{n-1}(x)  +  R_{n-2}(x) \; .
\end{equation}
Here, the remainder $R_{n-2}(x)$ is a polynomial in $x$ of {\em at most}
degree $n\!-\!2$, but usually of considerably smaller degree (though
we do not know how to quantify this at the moment).
The first two terms on the right hand side of (\ref{recpoly}) are
clear: one has $x\!-\!1$ possibilities to extend a square-free word
of length $n$ into one of length $n\!+\!1$ without a square of length
2 at the end. {}From these (if $n>2$) essentially
$P_{n-1}(x)$ words have to be subtracted because they contain
a square of length 4 at the end.
Further restrictions reach deaper into the word, as can be seen from
the structure of the stop-words.

Consequently, at least for large $x$, one would expect a reasonable 
approximation by simply neglecting the lower order terms. So, let us 
consider polynomials $Q_n(x)$ defined through
\begin{equation} \label{newpoly}
    Q_{n+1}(x) \; = \; (x-1) \cdot Q_n(x) - Q_{n-1}(x)
\end{equation}
with initial conditions $Q_0(x)\equiv 1$ and $Q_{-1}(x)\equiv 0$.
In this case, the generating function can easily be calculated
\begin{equation}
    F(x,t) \; := \;  \sum_{n=0}^{\infty} Q_n(x) \cdot t^n
              \; = \;  \frac{1}{t^2 - (x-1) t + 1}
\end{equation}
by applying the recurrence relation (\ref{newpoly})
and observing the initial conditions properly. 

For a given $x$, the growth rate of the coefficients (which we know
to converge) is given by the inverse of the radius of convergence $\varrho$
of the generating function wherefore we obtain the simple formula
\begin{equation} \label{guess1}
    \tilde{s}(x)  \; = \;  \log \left(
     \frac{(x-1) + \sqrt{(x-1)^2 - 4}}{2} \right) \, .
\end{equation}
Although the first two terms in the asymptotic expansion of (\ref{guess1})
match those of the upper bound, $\log(x\!-\!1) \sim \log(x) - x^{-1}$,
for finite $x$ it gives a much better approximation to the true value
of the entropy (see Table 3) except for $x=3$, where $\tilde{s}(3)=0$.
The lower bounds are those due to Proposition 1 with
$\varepsilon = \log(2)/21$ while the estimate is obtained from
counting square-free words in $x$ letters up to length $n_{\mbox{\tiny max}}$
(by extrapolating the logarithm of the successive ratios). The error
is roughly 1 figure in the last digit. The upper bound is again strict
and was calculated from Eq.~(\ref{entropy2a}) as
\begin{equation}
s_{}^-(x)  \; \leq \;  \frac{1}{n_{\mbox{\tiny max}}\!-2} \;\,
\log\left(\frac{\omega_{n_{\mbox{\tiny max}}}^-(x)}{x(x-1)}\right) \, .
\end{equation}

The convergence of $\tilde{s}(x)$ is rather quick (see Table~3), and the above
arguments indicate that $\tilde{s}(x)$ is asymptotically exact --
a property that deserves further investigation.

{\footnotesize
\vspace{10mm}
\centerline{Table 3: Bounds and estimates for the entropy of square-free
words in $x$ letters.}\samepage
\addtolength{\tabcolsep}{1mm}
\begin{center}
\begin{tabular}{|r|c|c|l|c|c|c|}
\hline
\multicolumn{1}{|c}{\rud $x$} &
\multicolumn{1}{|c}{$n_{\mbox{\tiny max}}$} &
\multicolumn{1}{|c}{lower bound} &
\multicolumn{1}{|c}{estimate} &
\multicolumn{1}{|c}{upper bound} &
\multicolumn{1}{|c}{$\log(x\!-\!1)$} &
\multicolumn{1}{|c|}{$\tilde{s}(x)$} \\
\hline\ru
 3 & 90 & 0.03300701 & 0.263719 & 0.27825962 & 0.69314718 & 0.00000000 \\
 4 & 26 & 0.26405607 & 0.96375  & 0.97319304 & 1.09861229 & 0.96242365 \\
 5 & 21 & 0.43734286 & 1.317089 & 1.32225469 & 1.38629436 & 1.31695790 \\
 6 & 18 & 0.57597230 & 1.56682  & 1.57028618 & 1.60943791 & 1.56679924 \\
 7 & 16 & 0.69149683 & 1.76275  & 1.76530829 & 1.79175947 & 1.76274717 \\
 8 & 16 & 0.79051786 & 1.924850 & 1.92663981 & 1.94591015 & 1.92484730 \\
 9 & 15 & 0.87716125 & 2.063438 & 2.06486642 & 2.07944154 & 2.06343707 \\
10 & 14 & 0.95417761 & 2.184644 & 2.18583786 & 2.19722458 & 2.18464379 \\
11 & 12 & 1.02349232 & 2.29243  & 2.29357100 & 2.30258509 & 2.29243167 \\ \rd
12 & 12 & 1.08650570 & 2.38953  & 2.39045454 & 2.39789527 & 2.38952643 \\
\hline
\end{tabular}
\end{center}
\addtolength{\tabcolsep}{-1mm}\vspace{10mm}
}

\section{Concluding remarks}

In this article, we have discussed various aspects of the ensemble
of square-free words in a finite alphabet with $x$ letters, with
some emphasis on the entropy density in the thermodynamic limit.
Though we could give various rigorous bounds for the entropy
(which is, in particular, strictly positive for $x>2$), we were
neither able to solve the problem analytically nor able to
construct an exhaustive lower bound (while this is easy for
the upper bound). Nevertheless, an approximate generating
function was given which results in an entropy estimate
that is asymptotically exact and astonishingly accurate already
for small $x>3$.

Several questions remain open. Although standard criteria point
to non-solvability of the problem (in the sense that the generating
function has the circle of convergence as its analyticity domain),
this needs further investigation. Here, a better understanding
of the lower bound would help because it would shed more light
onto possbile methods to essentially exhaust the square-free
words, at least w.r.t.~their exponential growth. We hope to
report on further findings in this direction in the near future.

\section*{Acknowledgements}
It is a pleasure to thank J.-P.~Allouche and
J.~Cassaigne for discussions and various useful comments.
M.~B.\ would like to thank B.~Iochum and the CPT Luminy
for  hospitality and financial support during a stay in spring 1995
where part of this work was done.
V.~E.\ thanks the Alexander von Humboldt Foundation for support.
U.~G.\ gratefully acknowledges financial support of the 
Samenwerkingverband FOM/SMC Mathematische Fysica.


\end{document}